\documentstyle[a4,amssymb,twoside,fleqn,espcrc2]{article}

\newcommand{\be}{\begin{equation}}
\newcommand{\ee}{\end{equation}}
\newcommand{\bea}{\begin{eqnarray}}
\newcommand{\eea}{\end{eqnarray}}
\title{Symmetries and solvable models for evaporating 
	2D black holes\thanks{Work partially supported by the {\it
	Comisi\'on Interministerial de Ciencia y Tecnolog\'{\i}a} 
	and {\it DGICYT}.}}

\author{J.~Cruz, J.~Navarro-Salas\address{Departamento de 
	F\'{\i}sica Te\'orica and IFIC,
	Centro Mixto Universidad de Valencia-CSIC,\\
	Facultad de F\'{\i}sica, Universidad de Valencia, Burjassot, 46100,
	Valencia, Spain.},
	M.~Navarro\address{Instituto de Matem\'aticas
	y F\'{\i}sica Fundamental,\\
	CSIC, Serrano 113-123, 28006 Madrid, Spain.}
	and C.~F.~Talavera\address{Departamento de Matem\'atica Aplicada, 
	E.T.S.I.I.,\\
	Universidad Polit\'ecnica de Valencia, 
	Camino de Vera, 46100, Valencia, Spain.}}

\begin{document}

\begin{abstract}
We study the evaporation process of a 2D black hole in thermal equilibrium
when the ingoing radiation is switched off suddenly.
We also introduce global symmetries of generic 2D dilaton gravity models
which generalize the extra symmetry of the CGHS model.
\end{abstract}

\maketitle

\section{INTRODUCTION}

In recent years it has been a lot of activity in the study of
two-dimensional dilaton gravity models.
The string-inspired model of Callan, Giddings, Harvey and Strominger 
\cite{CGHS} is the simplest model that describes the formation of black
holes by gravitational collapse.
The existence of an extra symmetry for the CGHS model plays a fundamental
role at the classical and quantum level.
The one-loop quantum corrected model of Russo, Susskind and Thorlacius
\cite{RST} was constructed to preserve the free field associated with the
extra symmetry and the classical ground state.
The model of Bose, Parker and Peleg \cite{BPP} also preserves the extra
symmetry but describes an evaporating black hole with a non-flat remnant
geometry.

The content of this work has two folds.
In Sections 2 and 3 we shall study the one-parameter 
family of models interpolating
between the BPP and RST models \cite{Cruz1} and the 
exponential model \cite{Mann}\cite{CNNT} respectively.
However, instead of studying the evaporation of a black hole formed by
gravitational collapse we shall analyze the semiclassical evaporation of a
black hole initially in thermal equilibrium when the incoming thermal flux
is switched off suddenly. 
In Section 4 we introduce global extra symmetries for generic 2D dilaton
gravity models, aiming to generalize the extra symmetry of the one-loop
corrected theories of the CGHS model.

\section{THERMAL BATH REMOVAL IN THE BPP-RST MODELS}

Let us consider the one-parameter family of models interpolating between the
BPP and RST models \cite{Cruz1},
\bea
 S &=& S_0 + S_P + \frac{N}{24\pi} \int d^2x \, \sqrt{-g} 
 \label{ec:i} \\
&& \qquad \left[(1-2\phi)(\nabla\phi)^2 + (a-1)\phi R \right] \, , 
\nonumber
\eea
where $S_0$ is the classical CGHS action
\bea
S_0 &=& \frac{1}{2\pi} \int d^2x \, \sqrt{-g} \Bigl[
e^{-2\phi} (R   \label{ec:ii} \\
&& \quad + 4(\nabla\phi)^2 + 4\lambda^2) - \frac{1}{2} \sum_{i=1}^N
(\nabla f_i)^2 \Bigr]  \, ,
\nonumber
\eea
$S_P$ is the Polyakov term and $a$ is an arbitrary real parameter.
In the conformal ($ds^2 = - e^{2\rho} dx^+ \, dx^-$) and Kruskal gauge
($\rho=\phi$) it is very easy to see that the above models admit the
following solution
\be
e^{-2\phi} + N\frac{a}{12}\phi = - \lambda^2 x^+ x^- + \frac{M}{\lambda}
\, ,
\label{ec:iii}
\ee
which represents a black hole in thermal equilibrium at temperature
$T=\frac{\lambda}{2\pi}$.
We can now remove the thermal bath by introducing a discontinuity in the
boundary condition \cite{Cruz2}
\bea
t_{x^+} &=& \frac{1}{4(x^+)^2} \theta(x^+ - x^+_0) \, , \label{ec:iv} \\
t_{x^-} &=& 0 \, . \label{ec:v}
\eea
In this way we get the dynamical solution (for $x^+ > x^+_0$)
\bea
e^{-2\phi} &+& N\frac{a}{12}\phi = -\lambda^2 x^+ (x^- + \Delta) 
\label{ec:vi} \\
&& \quad -\frac{N}{48} \left( \log\frac{x^+}{x^+_0} + 1 \right) 
 + \frac{M}{\lambda}
\, , 
\nonumber
\eea
where $\Delta = - \frac{N}{48 \lambda^2 x^+_0}$. 
It can be seen that the curvature singularity
 and the apparent horizon
intersect at the point
\bea
x^+_{\textrm{\small int}} &=& x^+_0
e^{\frac{4}{k}\left(\frac{M}{\lambda}-\alpha\right)} \, , \label{ec:vii} \\
x^+_{\textrm{\small int}} &=& \frac{k}{4 \lambda^2 x^+_0} \left[
1 - e^{-\frac{4}{k}\left(\frac{M}{\lambda} - \alpha\right)} \right] \, .
\label{ec:viii}
\eea
where 
\be
\alpha=N{a\over24}\left(1-\log N{a\over24}\right)\, .\label{ix}
\ee
The evaporating solution can be continuously matched to the static solution
\bea
e^{-2\phi} &+& N\frac{a}{12}\phi  = 
-\lambda^2 x^+ (x^- + \Delta) \label{ec:x} \\
&& \quad - \frac{k}{4} \log\left(-\lambda^2 x^+ (x^- +
\Delta)\right) + C \, ,
\nonumber
\eea
where 
\be
C = \frac{k}{4} \left( \log\frac{k}{4} - 1 \right) + \alpha \, ,
\label{ec:xi}
\ee
with the emission of a thunderpop at the intersection point with negative
energy $E_{\textrm{\tiny thunderpop}} = - \lambda\frac{k}{4}$,
independent of the initial state.
The remnant geometry (\ref{ec:x}) coincides with the one obtained in the
process originated by gravitational collapse.

\section{THERMAL BATH REMOVAL IN THE EXPONENTIAL MODEL}

The semiclassical theory of the exponential model \cite{Cruz2}
is given by the action 
\bea
S &=& {1\over2\pi}\int d^2x\sqrt{-g} \Bigl(
        \phi R+4\lambda^2e^{\beta\phi}
        \nonumber\\
&&\qquad -{1\over2}\sum_{i=1}^N (\nabla f_i)^2
\Bigr)
+S_P\nonumber\\
&& + {N\beta\over96\pi}
        \int d^2x\sqrt{-g}\left(\phi R+\beta\left(\nabla\phi\right)^2\right)
\, ,\label{ec:xii}
\eea
where $S_P$ is the Polyakov action.
The classical limit
\bea
S&=&{1\over2\pi}\int d^2x \, \sqrt{-g}
\Bigl(\phi R+4\lambda^2e^{\beta\phi} \nonumber\\
&&\qquad -{1\over2}\sum_{i=1}^N\left(\nabla f_i \Bigr)^2
\right)\, ,\label{ec:xiii}
\eea
arises as a theory with the conformal extra symmetry
\bea
\delta g_{\mu\nu}&=& -\epsilon \beta g_{\mu\nu} \, , \nonumber\\
 \delta\phi&=&\epsilon\, .\label{ec:xiv}
 \eea
The action (\ref{ec:xii}) contains the adequate counterterms 
in order to preserve
this symmetry when we add the Polyakov action, in analogy to the 
BPP-RST models.
In absence of matter and taking the boundary conditions $t_{\pm}=0$
we have the solution (in Kruskal gauge, $2\rho=\beta\phi$)
\be
ds^2=-\frac{dx^+dx^-}{\frac{\lambda^2\beta\gamma}{C}+Cx^+x^-}\, ,\label{ec:xv}
\ee
where $\gamma=1/\left( 1+{N\beta\over12} \right)$.
If $C<0$ (and $\lambda^2<0$) it represents a black hole
in thermal equilibrium with mass
$M={2\over\beta\gamma\pi}\sqrt{C}$ and temperature $T_H={\beta\gamma\over4}M$
proportional to its mass.
We now introduce the boundary conditions (\ref{ec:iv}),(\ref{ec:v})
and (in order to simplify calculations) the incoming shock-wave
\be
T_{++}={1\over2x_0^+\beta\gamma} \, \delta\left(x^+-x^+_0\right)
 \, .\label{ec:xvi}
\ee
With these ingredients and in the limit $N\beta>>1$, we get the solution
(for $x^+>x^+_0$)
\be
ds^2 = 
-\frac{\left(
{x_0^+\over x^+}
\right)^{1/2}}{{\lambda^2\beta\gamma\over C}+Cx^+_0x^-
\left(\log{x^+\over x^+_0}\right)} dx^+dx^-
\, .\label{ec:xvii}
\ee
One can check that the curvature singularity and the apparent horizon 
do not intersect each other at any finite point.
In contrast with the BPP-RST models the asymptotically flat 
coordinate $\sigma^+$ at past null infinity
 is not the same as for the initial solution implying that the incoming flux 
 has not been actually removed.
 In fact, it can be seen that (with an adequate choice of the point $x_0^+$)
 the incoming flux starts to decrease continuously at $x^+=x^+_0$,
vanishes at a certain time $x_1^+$ and for $x^+>x^+_1$
becomes a small negative flux which goes to zero at infinity. 
The incoming flux cannot affect qualitatively
the result because at late times it becomes negative and even so the 
black hole does not evaporate.

We notice that recently a procedure of thermal 
bath removal \cite{Cruz4} has been
applied to the Schwarzschild black hole by applying an operator 
formalism \cite{Mikovic} to quantize the Vaidya solution.

\section{EXTRA SYMMETRIES}

The action of the exactly solvable models (\ref{ec:i}) can be rewritten as
follows
\be
S = S_0\left(\bar{g}_{\mu\nu}, e^{-2\phi}+ \frac{N}{12} a e^{-2\phi}\right) +
S_P(\bar{g}_{\mu\nu}) \, ,
\label{ec:xviii}
\ee
where $\bar{g}_{\mu\nu} = g_{\mu\nu} e^{-2\phi}$ is an auxiliary metric
which is invariant under the symmetry transformation
\bea
\delta\phi &=& \epsilon \, e^{2\phi} \, , \label{ec:xix} \\
\delta g_{\mu\nu} &=& 2 \epsilon \, g_{\mu\nu} e^{2\phi} \, . \label{ec:xx}
\eea
This simple reconstruction of the one-loop quantum corrected CGHS models
suggest to investigate the existence of additional extra symmetries for
generic 2D dilaton models to construct solvable semiclassical models.

The general 2D dilaton gravity action can be written (through a conformal
reparametrization of the metric \cite{Banks,Gegenberg}) as:
\be
S = \frac{1}{2\pi} \int d^2x \, \sqrt{-g} \left( \phi R + V(\phi) \right) 
\, . \label{ec:xxi}
\ee
In conformal gauge this action can be identified with a 2D sigma-model
action \cite{Kazama,CNNT}
\be
S = \frac{1}{2\pi} \int d^2x \, \left( -4\partial_+ \phi \, \partial_- \rho +
\frac{1}{2} V(\phi) e^{2\rho} \right) \, ,
\label{ec:xxii}
\ee
where $(\rho,\phi)$ play the role of target-space coordinates and the
target-space metric is flat.
The condition for the existence of a symmetry in these models \cite{Kazama}
gives in this case
\be
\frac{d^2\log V(\phi)}{d\phi^2} = 0 \, , 
\label{ec:xxiii}
\ee
which has the general solution $V=4\lambda^2 e^{\beta\phi}$ with $\lambda$,
$\beta$ constants.
For $\beta=0$ we recover the CGHS model while for $\beta\neq 0$ we have 
the exponential model.

By direct computation one can check that the action (\ref{ec:xxi})
posseses the following extra symmetries for an arbitrary potential
\bea
\delta_1\phi &=& 0 \, , \nonumber \\
 \delta_1 g_{\mu\nu}&=&\epsilon_1
\left(\frac{g_{\mu\nu}}{\left(\nabla\phi\right)^2}-
2\frac{\nabla_{\mu}\phi\nabla_{\nu}\phi}{\left(\nabla\phi\right)^4}\right)
\, ,\label{ec:xxiv}
\eea
\bea
\delta_2\phi&=&\epsilon_2 \, ,\nonumber\\
\delta_2 g_{\mu\nu}&=&\epsilon_2 V
\left(\frac{g_{\mu\nu}}{\left(\nabla\phi\right)^2}-
2\frac{\nabla_{\mu}\phi\nabla_{\nu}\phi}{\left(\nabla\phi\right)^4}\right)
\, ,\label{ec:xxv}
\eea
\bea
\delta_3\phi&=&0 \, , \nonumber\\
 \delta_3g_{\mu\nu}&=&-\frac{\epsilon_3}{2}\Bigl[
g_{\mu\nu} \nonumber \\
&& + J
\left(\frac{g_{\mu\nu}}{\left(\nabla\phi\right)^2}
 -2\frac{\nabla_{\mu}\nabla_{\nu} \phi}
{\left(\nabla\phi\right)^4}\right)\Bigr]
\, .\label{ec:xxvi}
\eea
where ${dJ\over d\phi}=V(\phi)$,
which close down to a non-abelian Lie algebra in which $ \delta_2$
is a central generator, but $\delta_1$ and $\delta_3$ 
generate the affine subalgebra
\be
\left[\delta_1,\delta_3\right]={1\over2}\delta_1\, .\label{xxvii}
\ee
These three symmetries imply the existence of three free
field equations which are
\be
\square \int^{\phi}{d\tau\over 2E+J\left(\tau\right)}=0
\, ,\label{ec:xxviii}
\ee
\be
R+\square\log\left(\nabla\phi\right)^2=0
\, ,\label{ec:xxix}
\ee
\be
\square E=0
\, .\label{ec:xxx}
\ee
where $E={1\over2}\left(\left(\nabla\phi\right)^2-J\left(\phi\right)\right)$
is not only a free field but it is also a conserved quantity which can be
interpreted as the local energy of the solutions.

The symmetry
 $\delta=\delta_2-4\lambda^2\delta_1$ when $V=4\lambda^2$
 is just the conformal symmetry (\ref{ec:xix}),(\ref{ec:xx}) and
 therefore generalizes this symmetry for an arbitrary model.
 Moreover $\delta_{\beta}=\delta_2+2\beta\delta_3$ when
 $V=4\lambda^2e^{\beta\phi}$
 is exactly the conformal symmetry of the exponential model (\ref{ec:xiv}).
It is possible to construct an auxiliary metric 
\bea
\bar g_{\mu\nu}&=&\frac{2E_{\lambda}}{\left(\nabla\phi\right)^2}g_{\mu\nu}
\label{ec:xxxi} \\
&& \quad +\left(\frac{1}{2E_{\lambda}}
-\frac{2E_{\lambda}}{\left(\nabla\phi\right)^4}\right)
\nabla_{\mu}\phi\nabla_{\nu}\phi\, ,\nonumber
\eea
where $E_{\lambda}={1\over2}\left(\left(\nabla\phi\right)^2-J\left(\phi\right)
\right)+2\lambda^2\phi$, which is invariant under the symmetry $\delta$.
Relation (\ref{ec:xxxi}) can be easily inverted to give
\bea
g_{\mu\nu}&=&{2\bar E_{\lambda}\over\left(\bar\nabla\phi\right)^2}g_{\mu\nu}
\label{ec:xxxii} \\
&& \quad +\left(2\bar E_{\lambda} -2 {\bar E_{\lambda}\over
 \left(\bar\nabla\phi\right)^4}\right)\nabla_{\mu}\phi
\nabla_{\nu}\phi\, .\nonumber
\eea
With the aid of this metric we can construct a semiclassical action invariant
under
 $\delta$ by coupling the matter conformally to this invariant metric
\bea
S&=&S_{DG}\left(g\left(\bar g\right),\phi\right)\nonumber\\
&& -{1\over2}\sum_{i=1}^N\int d^2x\sqrt{-\bar g}
\bar g^{\mu\nu}\partial_{\mu}
f_i\partial_{\nu}f_i\nonumber\\
&&+S_P\left(\bar g\right)\, ,\label{ec:xxxiii}
\eea
where $S_{DG}$ is the action (\ref{ec:xxi}).
This action is invariant under the transformation
\bea
 \delta\phi&=&\epsilon \, , \nonumber\\
  \delta g_{\mu\nu}&=&0\, ,\label{ec:xxxiv}
 \eea
 and is the natural generalization of the BPP model for an arbitrary model of
 2D dilaton gravity.


\begin{thebibliography}{9}

\bibitem{CGHS} C.~G.~Callan, S.~B.~Giddings, J.~A.~Harvey and A.~Strominger,
Phys.~Rev.~D45 (1992) 1005.

\bibitem{RST} J.~G.~Russo, L.~Susskind and L.~Thorlacius, Phys.~Rev. D46
(1993) 3444; Phys.~Rev.~D47 (1993) 533.

\bibitem{BPP} S.~Bose, L.~Parker and Y.~Peleg, Phys. Rev. D52 (1995) 3512;
Phys. Rev. Lett. 76 (1996) 861; {\it Validity of the Semiclassical
Approximation and Back-Reaction}, WIS-MILW-96-th-10.

\bibitem{Cruz1} J.~Cruz and J.~Navarro-Salas, Phys.~Lett.~B 375 (1996) 47.

\bibitem{Mann}R.  B.  Mann, Nucl. Phys. B 418 (1994) 231

\bibitem{CNNT} J.~Cruz, J.~Navarro-Salas, M.~Navarro and C.~F.~Talavera,
{\it Symmetries and Black Holes in 2D Dilaton Gravity}, hep-th/9606097.

\bibitem{Cruz2} J.~Cruz and J.~Navarro-Salas, Phys.~Lett.~B 387 (1996) 51

\bibitem{Cruz4} J.  Cruz, A.  Mikovic and J.  Navarro-Salas,
{\it A quantum model for Schwarzschild black hole evaporation},
hep-th/9611219

\bibitem{Mikovic} A.  Mikovic and V.  Radovanovic,
{\it Two-Loop Back Reaction in 2D Dilaton Gravity},
hep-th/9606098 (to appear in Nucl. Phys. B).
 
\bibitem{Banks} T.~Banks and M.~O'Loughlin, Nucl. Phys. B362 (1991) 649.

\bibitem{Gegenberg} J.~Gegenberg, G.~Kunstatter and D.~Louis-Martinez,
Phys.~Lett.~B321 (1994) 193; Phys.~Rev.~D51 (1995) 1781; {\it Classical and
Quantum Mechanics of Black Holes in Generic 2D Dilaton Gravity}, 
gr-qc/9501017.

\bibitem{Kazama} Y.~Kazama, Y.~Satoh and A.~Tscuchiya, Phys.~Rev.~D51 (1995)
4265.

\end{thebibliography}
\end{document}